# Adaptive Control of Embedding Strength for Image Watermarking using Neural Networks

Mahnoosh Bagheri, Majid Mohrekesh, Nader Karimi, Shadrokh Samavi

*Isfahan University of Technology*

*Isfahan,* 84156-83111 *Iran*

*Abstract*— Digital image watermarking has been widely used in different applications such as copyright protection of digital media, such as audio, image, and video files. Two opposing criteria of robustness and transparency are the goals of watermarking methods. In this paper, we propose a framework for determining the appropriate embedding strength factor. The framework can use most DWT and DCT based blind watermarking approaches. We use Mask R-CNN on the COCO dataset to find a good strength factor for each sub-block. Experiments show that this method is robust against different attacks and has good transparency.

## I. INTRODUCTION

Digital image watermarking technology has been receiving more attention than before due to the rapid development of multimedia. This technology is a method for embedding a logo into the host image without causing a considerable visual change in the image. The logo could be copyright information, side information or any other information required to be embedded in the host image.

A watermarking method can be blind or non-blind. Non-blind methods, such as that of [1], need side information to extract watermark data but blind methods do not require such information. The embedding method can be in the spatial domain or the transform domain. The embedded watermark data should be robust enough against attacks and also transparent sufficient to meet the standard measures. Different methods have proposed during the past few years, for watermarking with different capabilities and restrictions. In [2] as an example, proper blocks are found for embedding in the host image, and a discrete cosine transform (DCT) is applied. Then coefficients that have fewer effects on the imperceptibility are selected, and information is embedded in those coefficients using a threshold. Margolis et al. [3] proposed a method that embeds in the transform domain. They use student-T distribution to find transform coefficients. DCT and discrete wavelet transform (DWT) are transform domains they use in their paper separately. Some papers cascade two transforms [4], [5] and [6]. The work done in [4] at first, applies DWT on image and then DCT on LL of the first level in the transformed image. Then it chooses the lower frequency coefficients of the final transformed image for embedding watermark string. In [5], after using ROI detection for selecting best host blocks, lowest ones in an ROI ranking, DCT is applied on sub-blocks of LH, HL, and HH in wavelet transform of them. Here like [2], the authors choose a couple of coefficients that have a lower effect on human eyes and also adequate robustness at the same time. Embedding phase in this paper finishes by conditional swapping of the coefficients couple considering each bit of watermark stream. The distance of couple value will be increased to enhance the robustness of their embedding method. We, in one of our previous works [6], used a cascade of DCT after DWT on a color space transforming. The embedding phase in this paper is different for each channel of YUV color space. As Y channel stores illuminance date, we embed in half of its capacity although the full capacity of U and V channels are used for embedding bits of watermark logo. In [7], after applying DWT on blocks of the image, an adaptive strength factor is used for embedding. The strength factor in this paper is computed by a function of wavelet energy and saliency of the image. Some research works used optimization algorithms to calculate better strength-factor, similar to [8]. The authors use the Cuckoo search algorithm in the wavelet domain to find the best strength factor. These mentioned papers used almost naïve methods without any serious use of artificial intelligence.

Around the year 2012, scientists started to use neural networks (NN) and similarly, researchers use these networks in watermarking methods. But most of the papers use neural networks for some parts of their methods. For example, an NN in [9] is used for finding the best strength factor. They used their own designed NN After applying a four-level wavelet transform on the host image. This NN results in predicting a sequence of coefficients that can choose the best strength factor for embedding. Zheng et al. [10] applied a DWT and then SVD on the host image and, after that, used a designed network. This network is trained to embed the watermark data in the transformed image. In [9] and [10], the authors both developed and trained a network for watermarking. As there is a lot of pre-trained networks, it seems reasonable to choose one of them and use it for embedding. Mask R-CNN [11] is one of the popular networks which is used for semantic segmentation. It was pre-trained on the COCO dataset [12] and can segment 48 classes in the target image. Mask R-CNN is an advanced version of Faster R-CNN [13]. Faster R-CNN is pre-trained on the PASCAL dataset to detect objects based on object locations hypothesis.

In this paper, we propose a blind method based on NN for calculating an adaptive strength factor. Our embedding phase consists of blocking, DWT, sub-blocking, and then DCT. The extraction phase is the inverse of the embedding phase. We also applied different attacks to embedded images to show the robustness of our proposed method.

This paper is organized as follows: The proposed method is given in Section II, the experimental results are shown in Section III, and the paper concludes in the last Section.

## II. Proposed Method

Our proposed watermarking method contains four main stages. In this section, we first explain a NN that segments the important parts of the image. Then, the way of computing strength factor and embedding phase is, and finally, extracting the logo image from the watermarked image is explained. Figure 1 shows our proposed scheme for adaptive image watermarking.

### A. Mask R-CNN Network

The new generation of Faster R-CNN is Mask R-CNN and is appropriate for semantic segmentation. The whole structure of Mask R-CNN is shown in Figure 2-a. The first stage of R-CNN finds candidates to be accepted as objects and points them by rectangular bounding boxes. This stage, called Region Proposal Network (RPN), is a neural network that uses a spatial sliding window to map images to a lower dimension vector. This vector is the input of two layers: the box-regression layer (reg layer) and the box-classification layer (cls layer). As shown in figure 2-b, if we have K regions, the cls layer and reg layer produce 2×K scores and 4×K coordinates, respectively [13]. The output of the second stage in Mask R-CNN is a mask that predicts the class and box offset for each ROI [11]. We used Mask R-CNN for our proposed method to get masks for each class in the host image.

### B. Compute Strength Factor

One of the outputs of Mask R-CNN contains masks for any given class. As a study, we defined a table that consists of two classes: person and car. We chose images inclusive of both person and car and extract their masks by Mask R-CNN. As Figure 3 shows, each mask multiplies by its coefficient that could be arbitrarily set considering any particular condition. The following equation produces $M_E$ as a final embedding map,

$$M_E = k_\alpha \left(1 - \max_{i \in \{C,P\}} (c_i \times M_i)\right) \quad (1)$$

where $C$ and $P$ represent for respective classes *Car* and *Person* and $c_i$ and $M_i$ are coefficients and maps for each class. $k_\alpha$ is a constant coefficient to scale $M_E$ for embedding. The number of classes could be more significant in the actual implementation of our method to cover any number of desired objects and ROIs. Mean of $M_E$ values corresponding to sub-block is computed as strength factor.

### C. Embedding Phase

The start stage of our embedding method is similar to our previous work [5] and applies a saliency detection on host grayscale image to locate blocks with minimum saliency as candidates of embedding. Figure 1 shows the overall operation of embedding information in the host image of our previous work [5], resize all images to 512×512. Then compute saliency

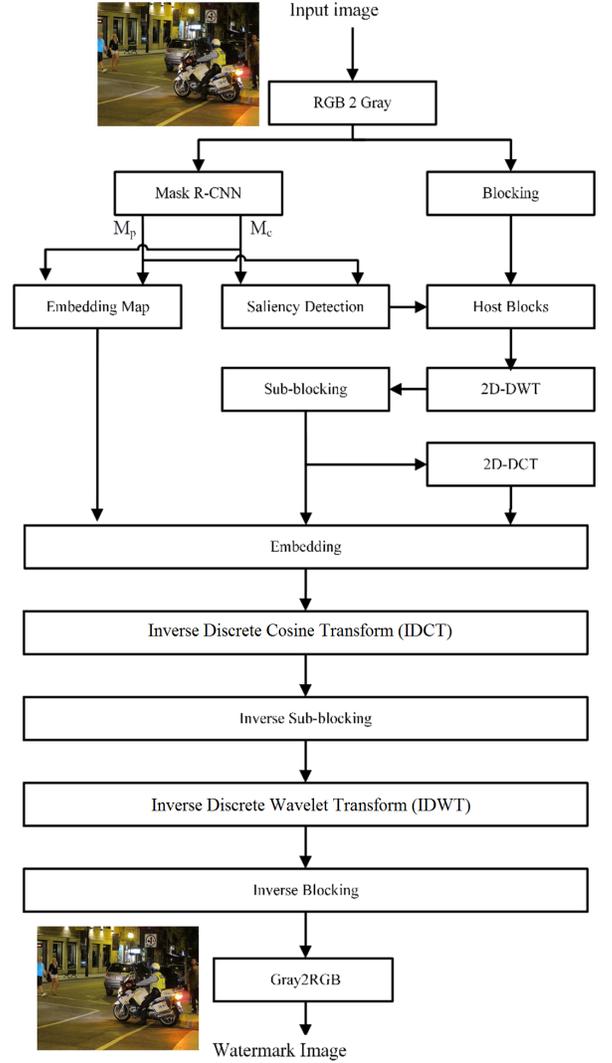

Fig 1. Block diagram of the proposed method.

for grayscale host images to find ROI areas by semantic segmentation from Mask R-CNN.

ROI areas have a lower rank in host image blocks and have a lower probability of embedding. Thus, after partitioning the host image to 128×128 blocks, a subset of them with lower ROI pixels are embedding candidates. The first five blocks of size 128×128 are selected, and two levels of DWT are applied to them, and then LH, HL, and HH parts will be partitioned to 8×8 sub-blocks. Then, we applied DCT on each sub-block and embed one bit in each 8×8 sub-block. Redundancy could be considered if the candidate sub-blocks were more than watermark bits.

The embedding process is done by the state of a couple of coefficients in the DCT domain. If the watermark bit is '1', DCT(6,7) should be higher than DCT(7,6) else, we swap them. For '0', the rule is inverse as equation 2.

$$\begin{cases} DCT\,(6,7) > DCT\,(7,6) & if\ w = 1 \\ DCT\,(6,7) < DCT\,(7,6) & if\ w = 0 \end{cases} \quad (2)$$

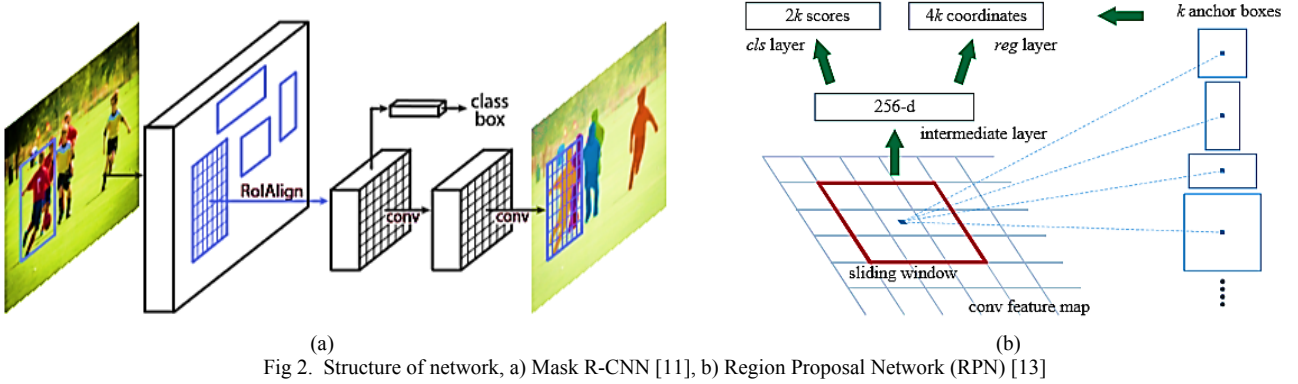

Fig 2. Structure of network, a) Mask R-CNN [11], b) Region Proposal Network (RPN) [13]

After that, we use a strength-factor that is calculated for each block. It enlarges the distance between the two coefficients to ensure maintaining the embedded bits in case of attacks to the host image.

We apply IDCT on each sub-block and transfer sub-blocks to 128×128 blocks after embedding watermark information.

Then we apply IDWT on each of five blocks and transfer them to create the watermarked image.

### D. Extracting Phase

The extraction of this method is blind and does not need the host image. When the watermarked image is delivered to the receiver, it will be partitioned to 128×128 blocks, and five selected blocks can be detected by indexes as side information. After that, DWT is applied to each of them and LH, HL, and HH parts are divided into 8×8 sub-blocks. Then, we use DCT on each, and the extraction method is based on Equation 3.

$$\begin{cases} 1 & if\ DCT\ (6,7) > DCT\ (7,6) \\ 0 & if\ DCT\ (6,7) < DCT\ (7,6) \end{cases} \quad (3)$$

We use a voting algorithm when redundant watermark bits are embedded in the same image to extract information that leverages the robustness of our method in the presence of attacks to the watermark image.

## III. EXPERIMENTAL RESULT

We test our method on images of the COCO dataset that has a person and car because our table concludes these classes. The COCO dataset has about 82K images that we select 93 images for evaluating our method. Input images have different sizes so at first, we resize all images to 512×512. The watermark bits by the size 4×4, one of the input images and it's watermarked image are shown in Figure 3. Our method is implemented in MATLAB R2015a in laptop Corei5-4200U CPU and 6GB of RAM and we use pre-train Mask R-CNN in Jupyter Notebook. The watermark information is a 4×4 logo that can be random that half of them should be 0, and the other half should be 1 like Figure 4-b.

### A. Evaluation Parameters

There are parameters that we can evaluate our methods by them. These parameters can show the goodness of our method and get the ability to compare our approach to other similar methods. One of them is the normalized correlation (NC)

$$NC = \frac{1}{M \times N} \sum_{i=1}^{M} \sum_{j=1}^{N} W(i,j) \times W'(I,j) \quad (4)$$

where $N$ and $M$ are sizes of the image, $W(I,j)$, and $W'(I,j)$ is watermark information and extracted watermark respectively. The other parameter is Structural Similarity (SSIM) where

$$SSIM = \frac{(2\mu i\ \mu j + c1)(2\sigma ij + c2)}{(\mu i^2 + \mu j^2 + c1)(\sigma i^{\wedge}2 + \sigma i^{\wedge}2 + c2)} \quad (5)$$

That μ and σ are mean and variance and $c_1$ and $c_2$ are constants. The Bit Error Rate (BER) is defined as below

$$BER = \frac{\sum_{i=1}^{M} \sum_{j=1}^{N} W(i,j) \oplus W'(I,j)}{M \times N} \quad (6)$$

The last parameter that we use in this paper is Peak Signal to Noise Ratio (PSNR), which is the influence of imperceptibility of an image.

$$PSNR = 20 \log_{10} \frac{256 \times 256}{MSE} \quad (7)$$

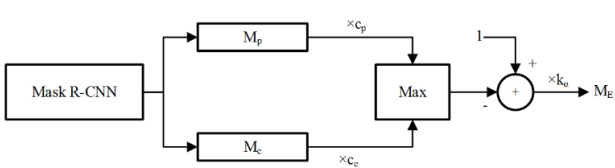

Fig 3. Embedding map formation.

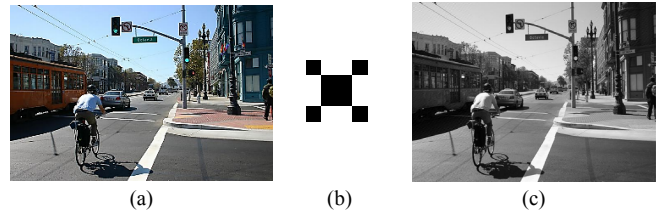

Fig 4. a) Input image, b) watermark bits, c) watermarked image

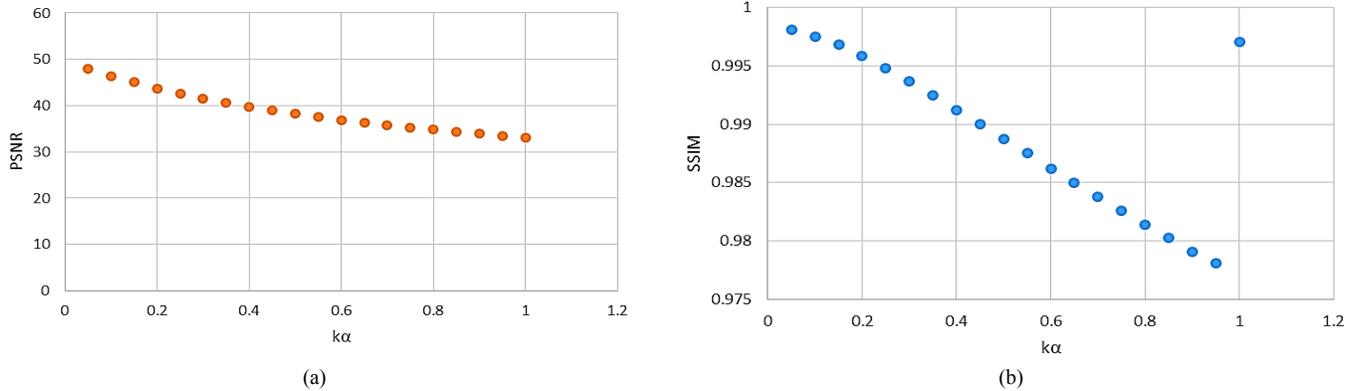

Fig 5. a) PSNR (dB), b) SSIM for different values of k.

## B. Results

We test our method for different parameters and compare them with [5]. The maximum PSNR and SSIM that our approach can achieve are 49.1052 and 0.9985 respectively. In comparison with [5], we have higher results. For example, in equal SSIM, we get 43.6043 for PSNR, but Jamali's PSNR is 42.4306. Results in Table 1 show our method is more robust against different attacks than [5]. Attacks that we test our approach by them are Gaussian noise (GN), JPEG compression (JC), 3×3 median filter (MF), histogram equalization (HE), and salt and pepper noise (S&P). We use three types of GN and JC that show the variance and ratio of compression respectively. Figures 5 show the effect of $k_\alpha$ on PSNR and SSIM that values of $c_p$ and $c_c$ are 1. By increasing $k_\alpha$ from 0.01 to 1, the value of PSNR and SSIM will be decreased. One of the results of our method represents in Figure 4.

## IV. CONCLUSION

In this paper, we proposed a method to use the neural network to find the strength factor of ROI based blind image watermarking. Each watermark bit embedded 15 times in the COCO dataset for redundancy improvement. Our method is robust against different attacks. In comparison to similar work, our approach has better results and in equal NC, has higher PSNR and SSIM.

Table 1. Results for different attacks

| Attack | | Proposed | | [5] | |
|---|---|---|---|---|---|
| | | BER | NC | BER | NC |
| GN | 0.01 | 0 | 1 | 0.0020 | 0.9960 |
| | 0.02 | 0 | 1 | 0.0040 | 0.9919 |
| | 0.03 | 0 | 1 | 0.0067 | 0.9866 |
| JC | 90 | 0 | 1 | 0.0074 | 0.9852 |
| | 80 | 0 | 1 | 0.0108 | 0.9785 |
| | 70 | 0 | 1 | 0.0128 | 0.9745 |
| MF | | 0 | 1 | 0 | 1 |
| HE | | 0 | 1 | 0 | 1 |
| S&P | | 0 | 1 | 0.0128 | 0.9745 |